\documentclass[]{pasj01}
\Received{$\langle$reception date$\rangle$}
\Accepted{$\langle$acception date$\rangle$}
\Published{$\langle$publication date$\rangle$}
\usepackage{bm}
\usepackage{scalefnt}
\usepackage{url}
\usepackage{color}
\newcommand{\red}{}{}
\newcommand{\redd}{}

\begin{document}

\title{Temporal and spectral X-ray properties of magnetar SGR\,1900+14 derived from observations with NuSTAR and XMM-Newton}
\author{Tsubasa \textsc{Tamba},\altaffilmark{1,}$^{*}$
Aya \textsc{Bamba},\altaffilmark{1,2}
Hirokazu \textsc{Odaka},\altaffilmark{1,2} and 
Teruaki \textsc{Enoto}\altaffilmark{3,4}}%
\altaffiltext{1}{Department of Physics, The University of Tokyo, 7-3-1 Hongo, Bunkyo-ku, Tokyo 113-0033, Japan}
\altaffiltext{2}{Research Center for the Early Universe, School of Science, The University of Tokyo, 7-3-1 Hongo,
Bunkyo-ku, Tokyo 113-0033, Japan}
\altaffiltext{3}{The Hakubi Center for Advanced Research, Kyoto University, Kyoto 606-8302, Japan}
\altaffiltext{4}{Department of Astronomy, Kyoto University, Kitashirakawa-Oiwake-cho, Sakyo-ku, Kyoto 606-8502, Japan}
\email{tamba@juno.phys.s.u-tokyo.ac.jp}


\maketitle

\begin{abstract}
X-ray observations play a crucial role in \red{understanding} the emission mechanism and relevant physical phenomena of magnetars. We report X-ray observations of a young magnetar SGR\,1900+14 made in 2016, which is famous for a giant flare \red{in} 1998 August. Simultaneous observations were conducted with XMM-Newton and NuSTAR on 2016 October 20 with 23 and 123 ks exposures, respectively. The NuSTAR hard X-ray coverage enabled us to detect the source up to 70 keV. The 1--10 keV and 15--60 keV fluxes were $3.11(3)\times10^{-12}\;{\rm erg\;s^{-1}\;cm^{-2}}$ and $6.8(3)\times10^{-12}\;{\rm erg\;s^{-1}\;cm^{-2}}$, respectively. The 1--70 keV spectra were well fitted by a blackbody plus power-law model with a surface temperature of $kT=0.52(2)\;{\rm keV}$, a photon index of the hard power-law of $\Gamma=1.21(6)$, and a column density of $N_{\rm H}=1.96(11)\times10^{22}\;{\rm cm^{-2}}$. Compared with previous observations with Suzaku in 2006 and 2009, the 1--10 keV flux showed a decrease by 25--40\%, while the spectral shape did not show any significant change with differences of $kT$ and $N_{\rm H}$ being within 10\% of each other. Through timing analysis, we found that the rotation period of SGR\,1900+14 on 2016 October 20 was $5.22669(3)\;{\rm s}$. The long-term evolution of the rotation period shows a monotonic decrease in the spin-down rate $\dot{P}$ lasting for more than 15 years. We also found a characteristic behavior of the hard-tail power-law component of SGR\,1900+14. The energy-dependent pulse profiles vary in morphology with a boundary of 10 keV. The phase-resolved spectra show the differences between photon indices ($\Gamma=1.02$--$1.44$) as a function of the pulse phase. Furthermore, the photon index is positively correlated with the X-ray flux of the hard power-law component, which could not be resolved by the previous hard X-ray observations.

\end{abstract}

\section{Introduction}
Soft gamma-ray repeaters (SGRs) and anomalous X-ray pulsars (AXPs) have recently been considered to form a class of young neutron stars with extremely strong magnetic fields \citep{Mereghetti1995, Kouveliotou1998}, which we call magnetars (\red{for a recent review, see \cite{Kaspi2017, Turolla2015}}). These objects exhibit rather slow rotation periods of $P=2$--$12\;{\rm s}$ and large spin-down rates of \red{$\dot{P}=10^{-15}$--$10^{-10}\;{\rm s\;s^{-1}}$} (\cite{Olausen2014}; McGill Online Magnetar Catalog\footnote{\url{http://www.physics.mcgill.ca/~pulsar/magnetar/main.html}}), which lead to huge dipole magnetic fields of $B_{\rm d}=10^{14}$--$10^{15}\;{\rm G}$, exceeding the quantum critical magnetic field $B_{\rm QED}=m_{\rm e}^{2}c^{3}/(e\hbar)=4.4\times10^{13}\;{\rm G}$ \citep{Harding2006}, where $m_{\rm e}$, $c$, $e$, and $\hbar$ denote the electron mass, speed of light, elementary charge, and Planck's constant, respectively. The radiation of magnetars is mainly emitted in the X-ray frequency and typically gives a luminosity of $10^{34}$--$10^{35}\;{\rm erg\;s^{-1}}$ \citep{Olausen2014}, which is much higher than the typical spin-down luminosity of magnetars of $10^{32}$--$10^{34}\;{\rm erg\;s^{-1}}$. \red{The small rotation power} compared to magnetar luminosity and the absence of evidence of accretion suggest that the magnetar is powered by liberating a part of its huge magnetic energy, but its mechanism is still unknown. X-ray observation of magnetars is a crucial step in \red{explaining} how they convert magnetic energy into radiation as well as what physical phenomena take place in their extremely strong magnetic fields (especially above $B_{\rm QED}$).\\
\indent It is widely known that magnetars typically show unstable fluctuations of spin-down rates $\dot{P}$ (e.g., CXOU J171405.7-381031: \cite{Halpern2010}; Swift\,J1822.3-1606: \cite{Tong2013}), while normal radio pulsars have constant $\dot{P}$ (e.g., the Crab pulsar: \cite{Terada2008}). The common mechanism of spin-down fluctuations of magnetars is still unclear. However, recent studies have suggested a common trend of the $\dot{P}$ fluctuations after experiencing outburst activities, which are also characteristic of magnetars. They seemed to show such unstable fluctuations of $\dot{P}$ followed by monotonic decreases (e.g., 1E\,1048.1-5937: \cite{Archibald2015}). A twisted magnetic field in the magnetosphere was proposed to explain the monotonic decreases in $\dot{P}$ by the decay of the twist \citep{Thompson2002, Beloborodov2009}. This seems to be a valuable common property of magnetars, but more samples are required to confirm it.\\
\indent Magnetars are also known for their hard-tail power-law components, which are dominant above $\sim10\;{\rm keV}$ with hard photon indices of $\Gamma\sim1$, coexisting with soft blackbody components with temperatures of $kT\sim0.5\;{\rm keV}$. Although the origin of the power-law component is still unknown, systematic studies on magnetar hard-tails have been conducted. They showed a trend that a magnetar with a younger characteristic age and stronger magnetic field displays a softer hard-tail \citep{Enoto2010, Enoto2017}. Photon splitting under extremely strong magnetic fields is a possible mechanism of the radiation \citep{Baring1998, Baring2001, Enoto2010}; however, this has not been confirmed owing to a lack of hard-tail observations.\\ 
\indent We require more magnetar samples with continuous broad-band X-ray monitoring to reveal these temporal and spectral properties. Compared to other magnetars, SGR\,1900+14 has a rather young characteristic age of $\tau_{\rm c}\sim0.9\;{\rm kyr}$ and a strong dipole magnetic field of $B_{\rm d}\sim7\times10^{14}\;{\rm G}$ \citep{Olausen2014}. It experienced a giant flare in 1998 August \citep{Hurley1999}, emitting a peak luminosity of $\gtrsim10^{44}\;{\rm erg\;s^{-1}}$ \citep{Mazets1999, Feroci2001}, which is much higher than ordinary outbursts of other magnetars \red{\citep{CotiZelati2018}}. Although it has been continuously observed for more than 20 years, the long-term variability of its rotation period after the giant flare has been poorly investigated. Its hard-tail power-law component was detected by INTEGRAL \citep{Gotz2006}, BeppoSAX \citep{Esposito2007} and Suzaku HXD \citep{Enoto2010, Enoto2017}, but they were not able to precisely determine its spectral and temporal properties because of large uncertainties. SGR\,1900+14 could be a valuable \red{resource} for studying common properties of magnetars if it is observed for a long period of time and with sufficient exposures.\\ %Therefore, we need a systematic analysis of the evolution of its rotation period to track the long-term variation, and also need an observation of SGR\,1900+14 with a sufficient statistics and a high sensitivity in the hard X-ray range to conduct a detailed analysis of the hard-tail power-law component.\\
\indent In this paper, we present analysis results of the simultaneous observations of SGR\,1900+14 with XMM-Newton and NuSTAR. The observation was conducted after sufficient time had passed \red{for tracking the long-term evolution} since the last observation in 2009. Making full use of the wide-band coverage of XMM-Newton and NuSTAR, we performed a detailed analysis, particularly on its hard-tail, for the first time. We also compare our results with previous ones and discuss the long-term evolution of SGR\,1900+14.

The remainder of this paper is organized as follows. In Section 2, we describe our observations and data reductions. Section 3 is devoted to the results of our observations. Then, we discuss our results in Section 4 and present our conclusions in Section 5.
%1. Overview of magnetars.\\
%2. Some details of magnetar properties that we focus on in this work.\\
%3. Historical background of SGR\,1900+14.\\
%4. What we have done in this paper.\\

\section{Observation and data reduction}
We observed SGR\,1900+14 simultaneously with XMM-Newton and NuSTAR on 2016 October 20. Table \ref{observationlog} shows details of the observations. Analyses were conducted using {\tt XSPEC} 12.10.0 and {\tt Xronos} 5.22.

\begin{table*}
  \tbl{ Observation logs. }{%
  \begin{tabular}{cccccc}
      \hline
      Telescope & ObsID & \begin{tabular}{c}Start Time\\(YYYY-MM-DD\\HH:MM:SS)\end{tabular} & \begin{tabular}{c}Stop Time\\(YYYY-MM-DD\\HH:MM:SS)\end{tabular} & \begin{tabular}{c}\red{Elapsed Time}\\(ks)\end{tabular} & \begin{tabular}{c}Net Exposure\\(ks)\end{tabular}\\ 
      \hline
      XMM-Newton & 0790610101 & 2016-10-20 21:43:05 & 2016-10-21 04:06:25 & 23.0 & \begin{tabular}{c}21.4 (MOS1, 2)\\ 11.3 (pn) \end{tabular} \\
      NuSTAR & 30201013002 & 2016-10-20 16:56:08 & 2016-10-23 12:01:14& 241.5 & 122.6 \\
      \hline
    \end{tabular}}\label{observationlog}
\begin{tabnote}
\end{tabnote}
\end{table*}

\subsection{XMM-Newton}\label{section:observation/XMM}
SGR\,1900+14 was observed for $23\;{\rm ks}$ with XMM-Newton, which is an X-ray telescope sensitive to 0.1--15 keV \citep{Jansen2001}; it carries three X-ray detectors: MOS1, MOS2, and pn \citep{Turner2001,Struder2001}. Throughout the observation, the MOSs were set in the large window mode, while pn was set in the full frame mode (time resolutions of $0.9\;{\rm s}$ and $73.4\;{\rm ms}$, respectively).\\
\indent All data were processed using XMM-Newton Science Analysis System (SAS) version 17.0.0, following the ``Users Guide to the XMM-Newton Science Analysis System''\footnote{\url{https://xmm-tools.cosmos.esa.int/external/xmm\_user\_support/documentation/sas\_usg/USG/}}. We omitted high background intervals by setting thresholds of 0.35 counts ${\rm s^{-1}}$ ($>$10 keV, single pixel events only) for the MOSs and 0.40 counts ${\rm s^{-1}}$ (10--12 keV, single pixel events only) for pn. As a result, the net exposure times became $21.4\;{\rm ks}$ for the MOSs and $11.3\;{\rm ks}$ for pn, as shown in Table \ref{observationlog}. We selected a circular source extraction region with a radius of $40\arcsec$ centered on SGR\,1900+14. For the MOSs, the background region was an annulus centered on the object with an inner radius of $40\arcsec$ and an outer radius of $144\arcsec$, while for pn, it was a circle with a radius of $108\arcsec$ at the source-free region on the same segment. We used {\tt rmfgen} and {\tt arfgen} in SAS to obtain the redistribution matrix files and ancillary response files, respectively. The spectra were re-binned by {\tt grppha} to have at least 50 counts in each bin. Barycentric corrected light curves were also generated using {\tt barycen} in SAS.

\subsection{NuSTAR}
\begin{figure*}
\begin{center}
\includegraphics[width=160mm]{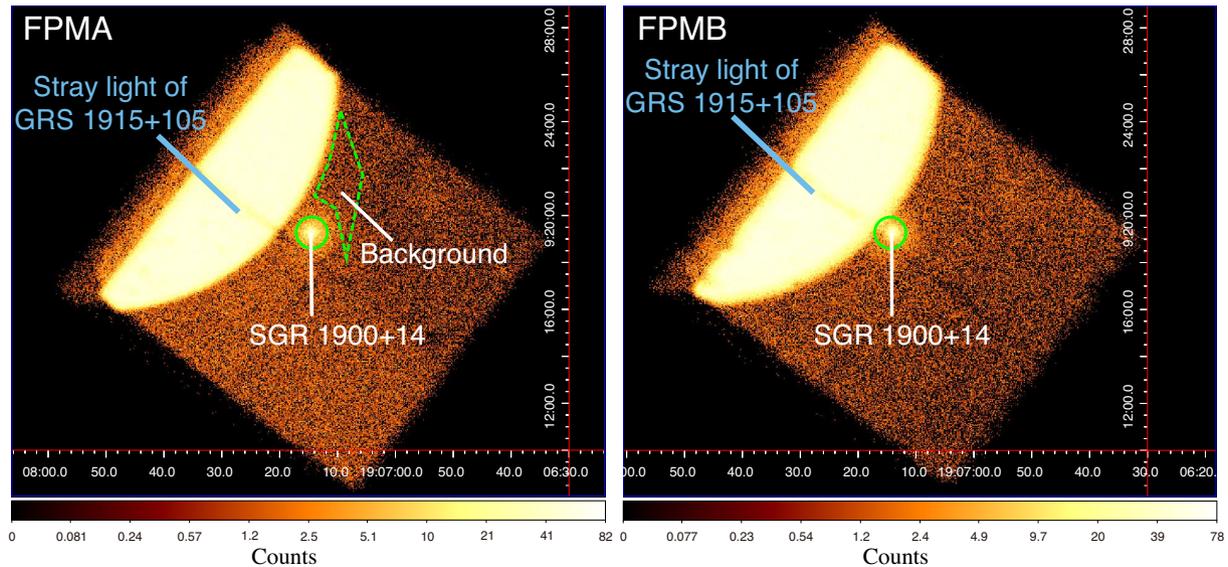}
\end{center}
\caption{NuSTAR 3--78 keV images for FPMA (left) and FPMB (right) in J2000 coordinates. The green circles denote the source region of SGR\,1900+14. The area surrounded by the green dashed line in the left image is the background region employed in this work. The extremely bright regions in both images are stray light from the nearby object GRS 1915+105.}
\label{NuSTARimage}
\end{figure*}

SGR\,1900+14 was observed for \red{an elapsed time of $242\;{\rm ks}$} with NuSTAR, which is the first focusing hard X-ray telescope covering 3--78 keV \citep{Harrison2013}. All data were processed using {\tt nupipeline} and {\tt nuproducts} in {\tt HEASoft} 6.23, following the ``NuSTAR Data Analysis Software Guide''\footnote{\url{https://heasarc.gsfc.nasa.gov/docs/nustar/analysis/nustar\_swguide.pdf}}. The net exposure time became 122.6 ks after the standard pipeline processes, as shown in Table \ref{observationlog}.\\
\indent Figure \ref{NuSTARimage} shows 3--78 keV images obtained from both detectors of NuSTAR, namely, FPMA and FPMB. The source region is the same as that in the XMM-Newton analysis. As shown in the upper left regions of both images, bright signals were caused by stray light from the nearby object GRS\,1915+105. Due to heavy contamination by the stray light in the source region, we decided not to use the FPMB data in this work. Other than GRS\,1915+105, there are two fainter sources of stray light called IGR\,J19140+0951 and 4U\,1908+075. Referring to the stray light simulation by the NuSTAR help desk\footnote{https://heasarc.gsfc.nasa.gov/cgi-bin/Feedback}, which showed background fluctuations, we selected the background region as shown in the left panel of Figure \ref{NuSTARimage} to avoid stray light contamination and to have the source and background region positioned in the same stray light area. We also checked the spectra generated with other background sets and found that there were no significant changes in the analysis. Setting the source and background region, we extracted spectrum and light curves. The spectrum was binned at minimum counts of 50 bin$^{-1}$ by {\tt grppha}, as described in \red{Section} \ref{section:observation/XMM}. The barycentric corrected light curve was also generated using {\tt barycorr} \red{by adopting the following coordinates for the source: RA=286.7891, DEC=9.3079}. %with parameters of {\tt srcra\_barycorr}=286.7891 and {\tt srcdec\_barycorr}=9.3079.

\section{Results}
\subsection{Spectral analysis}
Figures \ref{XMMspectrum} and \ref{NuSTARspectrum} show the XMM-Newton and NuSTAR spectra of SGR\,1900+14. These spectra are \red{featureless}, and \red{the NuSTAR spectrum extends up to 70 keV}. We conducted spectral fittings assuming a typical magnetar spectrum, a BB plus power-law (PL) \citep{Mereghetti2006, Enoto2010}. Photoelectric absorption was also taken into account. In {\tt XSPEC}, we employed a model {\tt phabs*(bbody+pegpwrlw)} to perform chi-squared fittings. \red{We employed the {\tt phabs} model with solar metallicity abundance angr \citep{Anders1989} and photoelectric absorption cross-section vern \citep{Verner1996}. We also tried another cross-section model, bcmc \citep{Balucinska-Church1992}, and confirmed no significant changes in our results.} The free parameters of the spectral fitting consist of hydrogen column density $N_{\rm H}$, BB surface temperature $kT$, BB normalization factor expressed in terms of luminosity, PL photon index $\Gamma$, and PL normalization factor in terms of the 2--10 keV unabsorbed flux. %Table \ref{bestfitparameters} shows summary of the spectral fittings. Errors indicate single-parameter 90\% confidence level.

\subsubsection{XMM-Newton}\label{section:spectrum/XMM}
\begin{figure*}
\begin{center}
\includegraphics[width=160mm]{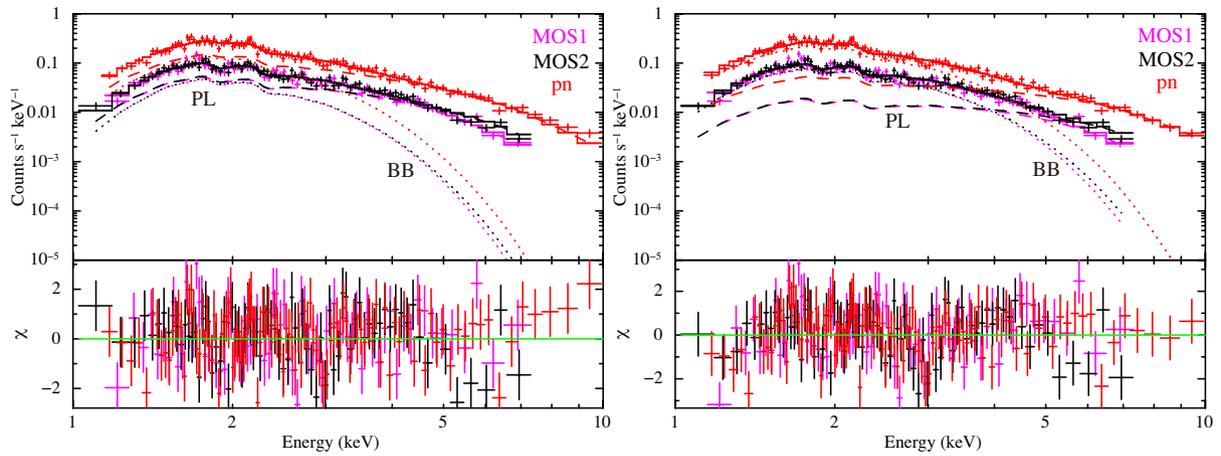}
\end{center}
\caption{XMM-Newton spectra fitted with blackbody (BB) + power-law (PL) model. $N_{\rm H}$ is set free in the left panel, while it is fixed to $1.9\times10^{22}\;{\rm cm^{-2}}$ in the right panel. Crosses are background-subtracted data and error bars represent $1\sigma$ confidence level. Each cross is binned with a minimum of 50 counts bin$^{-1}$. Magenta, black, and red crosses and lines represent results for MOS1, MOS2, and pn, respectively. Dotted, dashed, and solid lines are BB component, PL component, and the aggregation of the two, respectively.}
\label{XMMspectrum}
\end{figure*}

%We detected X-ray emissions from SGR\,1900+14 in the range of 0.1--10 keV for MOSs and 0.5--10 keV for pn. 
When conducting the spectral fitting, we omitted data below $1\;{\rm keV}$ due to poor statistics. We also discarded data above $8\;{\rm keV}$ for the MOSs due to poor statistics. As a result, the energy ranges for the fitting were 1--8 keV and 1--10 keV for the MOSs and pn, respectively.\\
\indent  The fitting returned a good reduced chi-squared $\chi_{\nu}^{2}$ (d.o.f.) of 1.08 (256) without large residuals. The best-fit model is presented in the left panel of Figure \ref{XMMspectrum} and the first row of Table \ref{bestfitparameters}. When all the five parameters are set free, the obtained \red{$N_{\rm H}=(2.6\pm0.3)\times10^{22}\;{\rm cm^{-2}}$} is significantly different from those obtained in studies by Suzaku, which were $(1.8\pm0.3)\times10^{22}\;{\rm cm^{-2}}$ and $(1.9\pm0.1)\times10^{22}\;{\rm cm^{-2}}$ \citep{Enoto2017}. This could be due to the absence of data above 10 keV, which leads to a failure in determining the photon index and thus $N_{\rm H}$. We checked the correlation contour of $N_{\rm H}$ and $kT$, and confirmed that the two parameters are significantly coupled. We thus conducted another spectral fitting with a fixed $N_{\rm H}$ of $1.9\times10^{22}\;{\rm cm^{-2}}$ in accordance with the previous studies by Suzaku \citep{Enoto2017}. This fitting yielded a similar acceptable $\chi_{\nu}^{2}$ (d.o.f.) of 1.15 (257) without large residuals. This best-fit model is shown in the right panel of Figure \ref{XMMspectrum} and the second row of Table \ref{bestfitparameters}.

\subsubsection{NuSTAR}\label{section:spectrum/NuSTAR}
\begin{figure}
\begin{center}
\includegraphics[width=80mm]{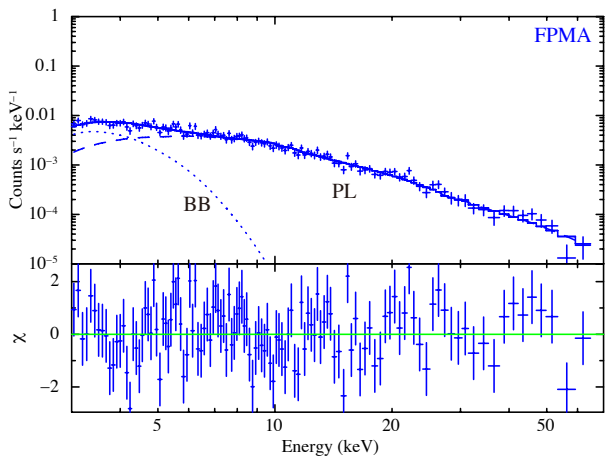}
\end{center}
\caption{NuSTAR FPMA spectrum fitted with blackbody (BB) + power-law (PL) model. Crosses are background-subtracted data and error bars represent $1\sigma$ confidence level. Each cross is binned with a minimum of 50 counts bin$^{-1}$. Dotted, dashed, and solid lines are BB component, PL component, and the aggregation of the two, respectively.}
\label{NuSTARspectrum}
\end{figure}

We detected X-ray emissions from SGR\,1900+14 in the energy range of 3--70 keV. The signal significance was $6.5\sigma$ in the range of 60--70 keV, while the Suzaku observation in 2006 detected the source only up to 50 keV \citep{Enoto2010}. Our observation realizes the first detection of SGR\,1900+14 above 50 keV after the detections by INTEGRAL in 2003 and 2004 \citep{Gotz2006}. In the fitting to NuSTAR data, $N_{\rm H}$ was fixed to $1.9\times10^{22}\;{\rm cm^{-2}}$, which was reported in the previous Suzaku studies \citep{Enoto2017}, because the photoelectric absorption does not have significant influence on the spectrum above 3 keV.\\
\indent The fitting applied to 3--70 keV yielded a good $\chi_{\nu}^{2}$ (d.o.f.) of 1.17 (118) without any distinctive structure in the residuals. Figure \ref{NuSTARspectrum} and the third row of Table \ref{bestfitparameters} describe the best-fit model. The obtained parameters are roughly consistent with those yielded with XMM-Newton (\red{section} \ref{section:spectrum/XMM}), but $\Gamma$ was determined more precisely due to the hard X-ray coverage of NuSTAR.

\subsubsection{Joint fitting}\label{section:spectrum/combined}
\begin{figure}
\begin{center}
\includegraphics[width=80mm]{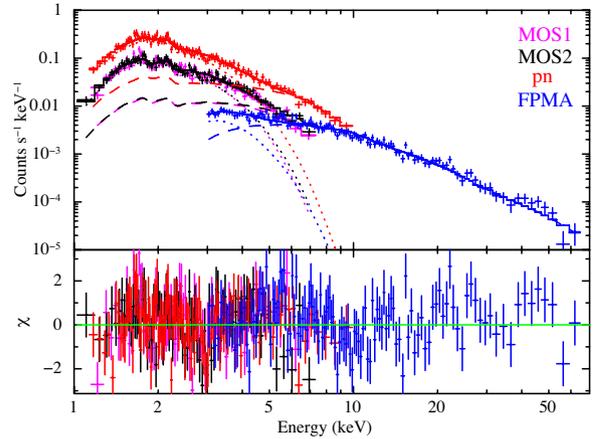}
\end{center}

\caption{XMM-Newton + NuSTAR spectra fitted with blackbody (BB) + power-law (PL) model. Crosses are background-subtracted data and error bars represent $1\sigma$ confidence level. Each cross is binned with a minimum of 50 counts bin$^{-1}$. Magenta, black, red, and blue crosses and lines represent results for MOS1, MOS2, pn, and FPMA, respectively. Dotted, dashed, and solid lines are BB component, PL component, and the aggregation of the two, respectively.}
\label{combinedspectra}
\end{figure}

We fitted the spectra of both XMM-Newton and NuSTAR simultaneously. We used the same energy ranges employed for each detector in the separate analyses (see \red{Sections} \ref{section:spectrum/XMM} and \ref{section:spectrum/NuSTAR}). \red{Although inter-calibration uncertainties between different instruments exist (e.g., see \cite{Tsujimoto2011}), the inter-calibration uncertainties between XMM-Newton and NuSTAR is up to 10\% \citep{Madsen2017}, and we fixed the cross normalization to 1 because it does not affect our results.} The parameter $N_{\rm H}$ was set free in this fitting. The fitting yielded an acceptable $\chi_{\nu}^{2}$ (d.o.f.) of 1.18 \red{(378)}. Although the residuals may show a distinctive structure, this does not affect the results significantly. The best-fit model is presented in Figure \ref{combinedspectra} and the fourth row of Table \ref{bestfitparameters}. \red{The absorbed 1--70 keV flux was $(1.21\pm0.04)\times10^{-11}\;{\rm erg\;s^{-1}\;cm^{-2}}$, where the error denotes 1$\sigma$ confidence level.} Owing to the wide-band spectral fitting, parameters $kT$ and $\Gamma$ were both successfully determined precisely. 

\begin{table*}
  \tbl{ Best-fit parameters of the spectral analysis.\footnotemark[$*$] }{%
  \begin{tabular}{ccccccc}
      \hline
      Telescope & \begin{tabular}{c}$N_{\rm H}$\\$(10^{22}\;{\rm cm^{-2}})$\end{tabular} &  \begin{tabular}{c}$kT$\\(keV)\end{tabular} & BB norm\footnotemark[$\dag$] & $\Gamma$ & PL norm\footnotemark[$\ddag$] & $\chi^{2}_{\nu}$ (d.o.f.)\\ 
      \hline
      XMM-Newton & \red{$2.6\pm0.3$} & $0.42^{+0.05}_{-0.06}$ & \red{$4.0^{+0.8}_{-1.0}$} & \red{$2.5\pm0.4$} & \red{$2.7\pm0.5$} & 1.08 (256)\\
      XMM-Newton & 1.9 (fixed)\footnotemark[$\S$] & $0.52^{+0.02}_{-0.01}$ & \red{$4.4\pm0.4$} & \red{$1.4\pm0.3$} & $1.78\pm0.14$ & 1.15 (257)\\
      NuSTAR & $1.9$ (fixed)\footnotemark[$\S$] & $0.64\pm0.07$ & \red{$3.1^{+0.8}_{-0.6}$} & $1.13\pm0.08$ & $1.65\pm0.13$ & 1.17 (118)\\
      XMM-Newton + NuSTAR & $1.96\pm0.11$ & $0.52\pm0.02$ & \red{$4.8\pm0.3$} & $1.21\pm0.06$ & $1.78\pm0.09$ & 1.18 (378)\\
      \hline
    \end{tabular}}\label{bestfitparameters}
\begin{tabnote}
\footnotemark[$*$] Errors denote single-parameter 90\% confidence level.  \\ 
\footnotemark[$\dag$] Normalization of the blackbody model is determined by the X-ray luminosity in units of $10^{34}\;{\rm erg\;s^{-1}}$ when assuming its distance at 10 kpc. Note that the latest estimation of the distance to SGR\,1900+14 is $12.5\pm1.7\;{\rm kpc}$ \citep{Davis2009}.\\
\footnotemark[$\ddag$] Normalization of the power-law model is determined by 2--10$\;{\rm keV}$ unabsorbed PL flux in units of $10^{-12}\;{\rm erg\;s^{-1}\;cm^{-2}}$.\\
\footnotemark[$\S$] This value is adopted in accordance with previous Suzaku studies \citep{Enoto2017}.
\end{tabnote}
\end{table*}

\subsection{Timing analysis}

\subsubsection{Time variability}
\begin{figure}
\begin{center}
\includegraphics[width=80mm]{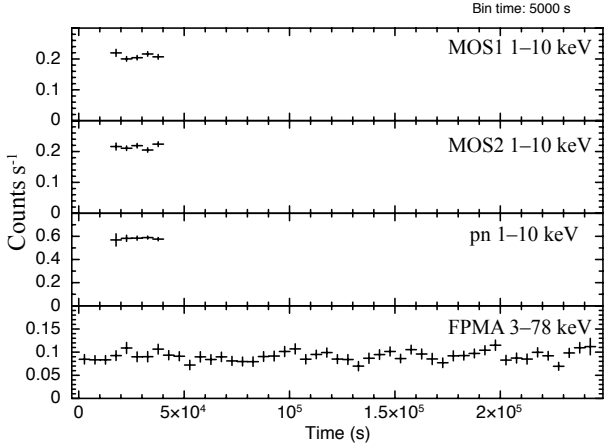}
\end{center}
\caption{Background-subtracted light curves of MOS1, MOS2, pn, and FPMA. The energy ranges employed are 1--10 keV for the MOSs and pn and 3--78 keV for FPMA. Each light curve is binned at $5000\;{\rm s\;bin^{-1}}$. Errors indicate $1\sigma$ confidence level.}
\label{lightcurve}
\end{figure}

Figure \ref{lightcurve} shows the light curves obtained from the observations. Chi-squared tests against being constant were conducted for each light curve, where the bin time was set to $1000\;{\rm s}$ for XMM-Newton and $5000\;{\rm s}$ for NuSTAR. For MOS1, MOS2, pn, and FPMA, the $\chi^{2}_{\nu}$ values are $1.22\;({\rm d.o.f.}=21)$, $0.99\;({\rm d.o.f.}=21)$, $0.50\;(\rm{d.o.f.}=15)$, and $1.02\;({\rm d.o.f.}=48)$, respectively, indicating no significant time variability in the XMM-Newton and NuSTAR data at the timescales of the binning times. \red{We also tried a smaller bin of $1\;{\rm s}$ and $0.1\;{\rm s}$ to the NuSTAR data and $1\;{\rm s}$ to the aggregation of the XMM-Newton data, and confirmed that there were no short bursts during the observation.}

\subsubsection{Coherent pulsation}\label{section:timing/coherentpulsation}
\begin{figure}
\begin{center}
\includegraphics[width=80mm]{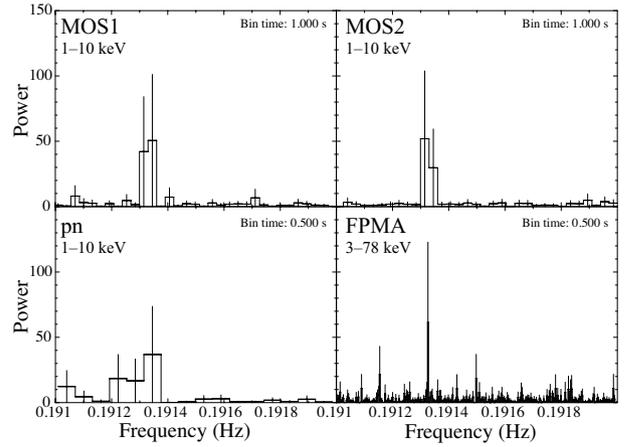}
\end{center}
\caption{Power spectra of MOS1, MOS2, pn, and FPMA. The energy ranges employed are 1--10 keV for the MOSs and pn, and 3--78 keV for FPMA. Binning time is $1.0\;{\rm s}$ for MOS1 and MOS2 and $0.5\;{\rm s}$ for pn and FPMA. Errors indicate $1\sigma$ confidence level.}
\label{powspec}
\end{figure}

Figure \ref{powspec} presents the power spectra of SGR\,1900+14 generated by applying Fourier transforms to the light curves with {\tt Xronos}. All the four power spectra show peaks at similar frequencies of $0.1913\;{\rm Hz}$ to $0.1914\;{\rm Hz}$. \red{In addition, we performed an epoch-folding search on the NuSTAR data.} As a result, we found a coherent pulsation of SGR\,1900+14 at a period of $5.22669\pm0.00003\;{\rm s}$ on 2016 October 20 to 23. Note that we were not able to detect any change of the pulsation period within this observation and conducted the epoch-folding search under the assumption that $\dot{P}=0$. We also confirmed that the results do not change significantly when adopting an appropriate value of $\dot{P}=10^{-10}\;{\rm s\;s^{-1}}$, \red{which is roughly equal to one, as derived by a previous XMM-Newton study with the same source \citep{Mereghetti2006}}.

\subsubsection{Pulse profiles}\label{section:timing/pulseprofiles}
\begin{figure}
\begin{center}
\includegraphics[width=80mm]{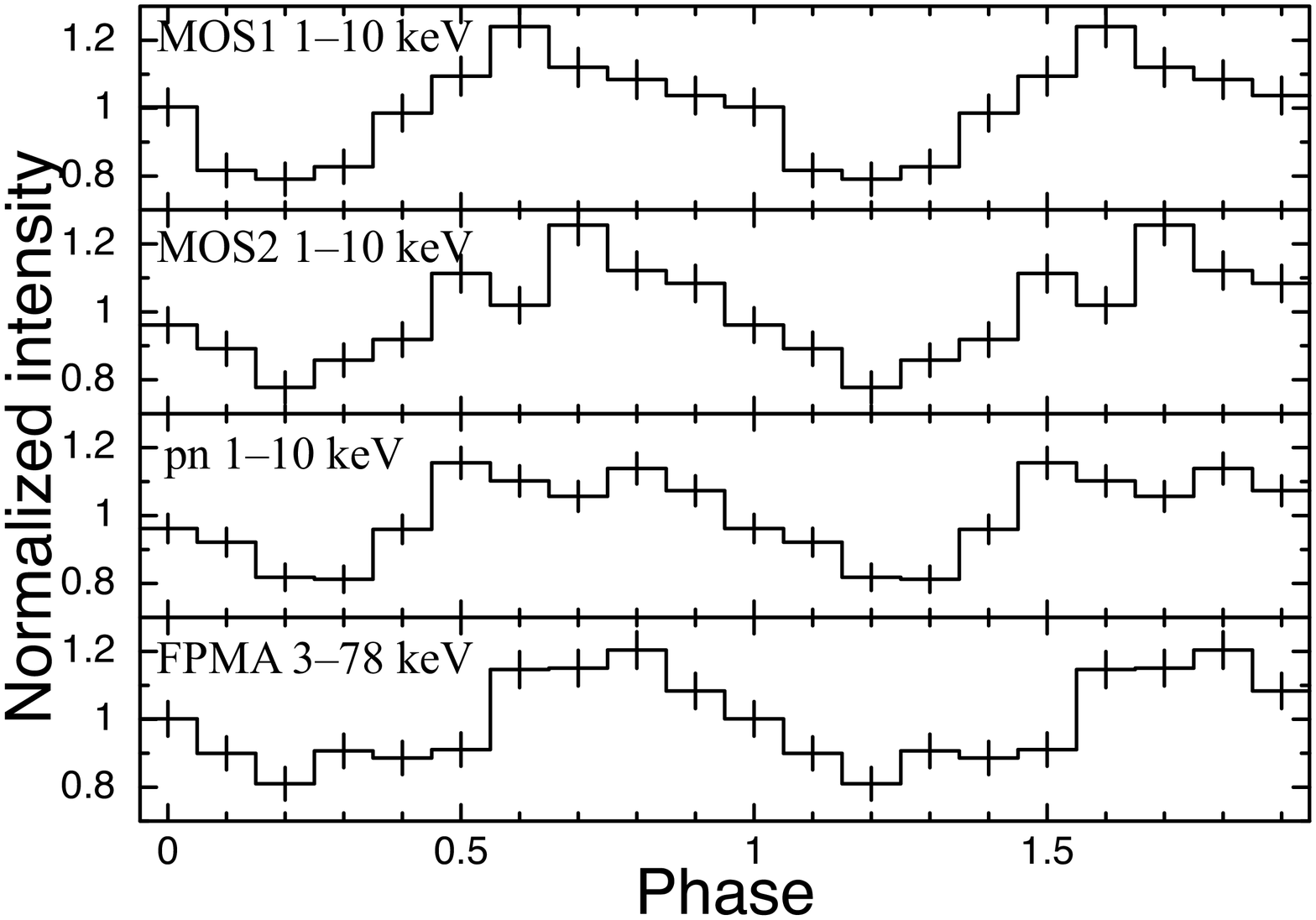}
\end{center}
\caption{Pulse profiles of MOS1, MOS2, pn, and FPMA folded by the rotation period $5.22669\;{\rm s}$. \red{Background is} subtracted. The energy ranges employed are 1--10 keV for the MOSs and pn and 3--78 keV for FPMA. The horizontal axis represents two cycles of pulsation. Errors indicate $1\sigma$ confidence level.}
\label{pulseprofiles}
\end{figure}

\begin{figure*}
\begin{center}
\includegraphics[width=160mm]{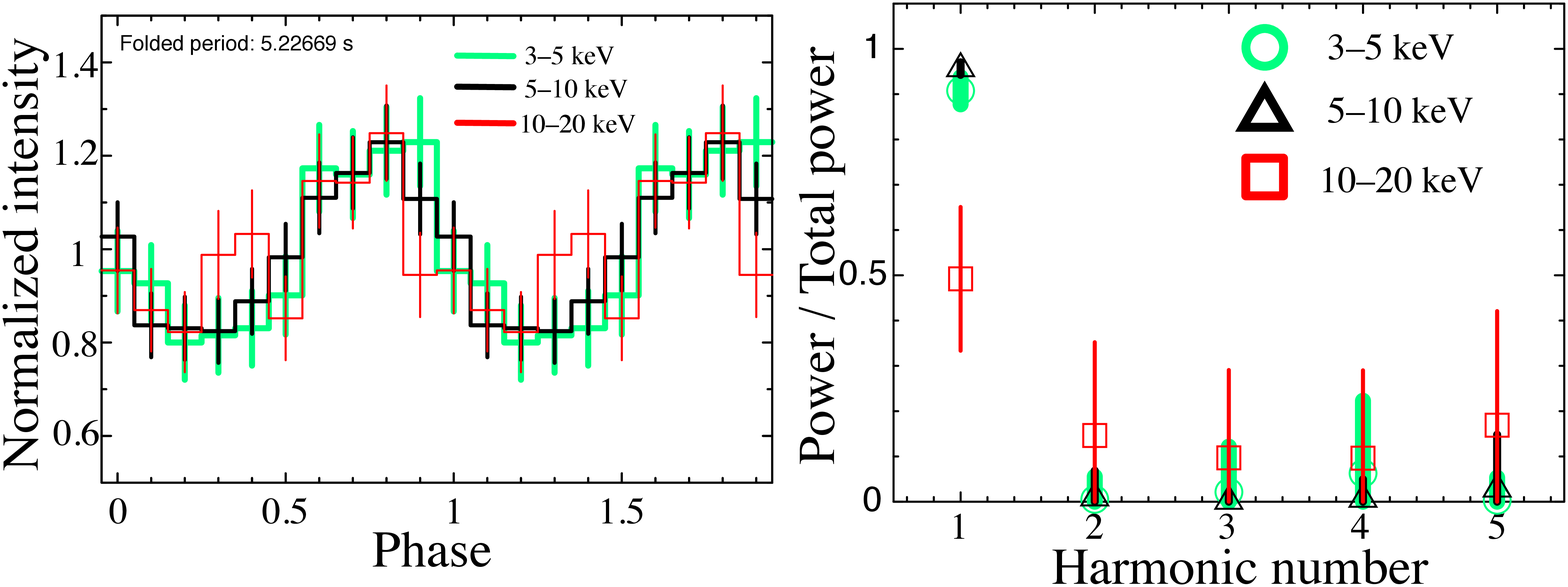}
\end{center}
\caption{Pulse profiles of FPMA 3--5, 5--10, and 10--20 keV (left). Fourier transforms of each pulse profile (right). \red{Background is} subtracted. Green, black, and red represent 3--5, 5--10, and 10--20 keV, respectively. Errors indicate $1\sigma$ confidence level.}
\label{pulseprofile_energy}
\end{figure*}

Here, we define the strength of pulsations by two different methods with reference to \citet{Tendulkar2015}. First, we define the rms pulse fraction as
\begin{eqnarray}
{\rm PF_{rms}}=\frac{2\sqrt{\sum_{k=1}^{k_{\rm max}}\left(\left(a_{k}^{2}+b_{k}^{2}\right)-\left(\sigma_{a_{k}}^{2}+\sigma_{b_{k}}^{2}\right)\right)}}{a_{0}}, 
\label{PFrms}
\end{eqnarray}
where $a_{k}$ and $b_{k}$ are Fourier coefficients expressed by
\begin{eqnarray}
a_{k}&=&\frac{1}{N}\sum_{j=1}^{N}p_{j}\cos\left(\frac{2\pi kj}{N}\right) \label{ak}\\
b_{k}&=&\frac{1}{N}\sum_{j=1}^{N}p_{j}\sin\left(\frac{2\pi kj}{N}\right),
\label{bk}
\end{eqnarray}
and $\sigma_{a_{k}}^{2}$ and $\sigma_{b_{k}}^{2}$ are the uncertainties in $a_{k}$ and $b_{k}$, respectively, expressed by 
\begin{eqnarray}
\sigma_{a_{k}}^{2}&=&\frac{1}{N^{2}}\sum_{j=1}^{N}\sigma_{p_{j}}^{2}\cos^{2}\left(\frac{2\pi kj}{N}\right) \label{sigmaak}\\
\sigma_{b_{k}}^{2}&=&\frac{1}{N^{2}}\sum_{j=1}^{N}\sigma_{p_{j}}^{2}\sin^{2}\left(\frac{2\pi kj}{N}\right). \label{sigmabk}
\end{eqnarray}
$N$, $p_{j}$, and $\sigma_{p_{j}}$ denote the number of phase bins, number of photons in each phase bin, and the Poisson variance in each phase bin, respectively. Note that the definition of ${\rm PF_{rms}}$ is modified from \citet{Tendulkar2015} so that ${\rm PF_{rms}}=B/A$ when the input signal is $A+B\sin\phi$ (see Appendix 1 of \citet{An2015} for more details). In this work, we set $k_{\rm max}=5$.\\
\indent We also employ the area pulse fraction ${\rm PF_{area}}$ described by
\begin{eqnarray}
{\rm PF_{area}}=\frac{\sum_{j=1}^{N}p_{j}-N*{\rm min}(p_{j})}{\sum_{j=1}^{N}p_{j}}, \label{PFarea}
\end{eqnarray}
which is defined in \citet{Gonzalez2010}.\\
\indent Figure \ref{pulseprofiles} shows the pulse profiles of all four detectors, which are folded by the best-fit pulsation period $5.22669\;{\rm s}$. All four profiles show their maxima at a phase of 0.6--0.8 and minima at a phase of 0.1--0.3. Although we did not consider the change of pulsation period $\dot{P}$, we have checked that the results do not change significantly when it is considered. We calculated the pulse fractions of all four profiles by employing the above equations ((\ref{PFrms})--(\ref{sigmabk}) and (\ref{PFarea})). The ${\rm PF_{rms}}$ (${\rm PF_{area}}$) values measured by MOS1, MOS2, pn, and FPMA are $18.6\pm2.6\%\;(20.9\pm4.9\%)$, $19.0\pm2.7\%\;(22.3\pm4.7\%)$, $15.6\pm2.0\%\;(18.7\pm4.0\%)$, and $17.0\pm2.5\%\;(19.0\pm5.0\%)$, respectively, with the error denoting $1\sigma$ confidence.\\% All the profiles show the consistent pulse fractions within their error ranges.\\
\indent We made energy-selected pulse profiles and conducted chi-squared tests against being constant. We found signals of periodic fluctuations below 20 keV with a confidence level of $99\%$, while we were not able to detect any pulse above 20 keV, presumably owing to poor statistics. The left panel of Figure \ref{pulseprofile_energy} shows NuSTAR pulse profiles of 3--5, 5--10, and 10--20 keV. This suggests that the pulse shape differs below and above 10 keV. To quantify this feature, we applied Fourier transforms to the pulse profiles, as shown in the right panel of Figure \ref{pulseprofile_energy}. The $i$-th harmonic power relative to the total power is calculated as
\begin{eqnarray}
\frac{P_{i}}{P_{\rm total}}=\frac{a_{i}^{2}+b_{i}^{2}}{\sum_{k=1}^{k_{\rm max}}(a_{k}^{2}+b_{k}^{2})}.\label{relativepower}
\end{eqnarray}
While the 3--5 keV and 5--10 keV profiles display almost sinusoidal profiles with the fundamental frequency dominantly contributing to the whole powers, the 10--20 keV profile shows contributions from higher harmonics. This result clearly indicates a change of the pulse profile at $\sim10\;{\rm keV}$. Because the power-law (PL) component is prominent above 10 keV (see Figure \ref{combinedspectra}), the appearance of the higher harmonics can be interpreted as a characteristic behavior of the PL component, while the blackbody component seems to have the sinusoidal profile.

\subsection{Phase-resolved spectra}\label{section:phasespectra}
\begin{figure*}
\begin{center}
\includegraphics[width=160mm]{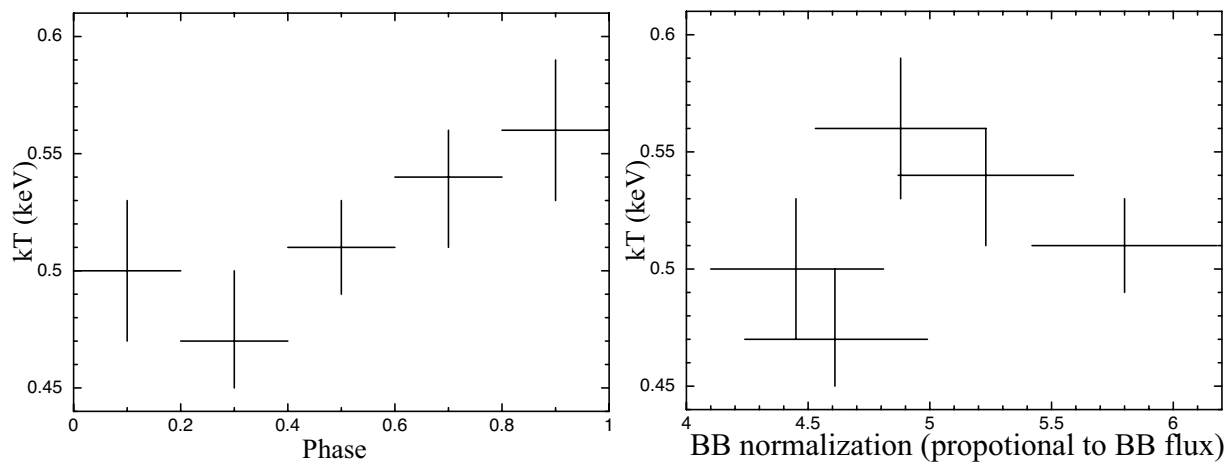}
\end{center}
\caption{(left) BB temperature versus pulse phase. The photon index of the PL component is fixed in the fitting. The vertical error bars denote single-parameter 90\% confidence level. (right) BB temperature versus BB normalization, which is proportional to BB flux. Both vertical and horizontal error bars denote single-parameter 90\% confidence level.}
\label{BBtempplot}
\end{figure*}

\begin{figure*}
\begin{center}
\includegraphics[width=160mm]{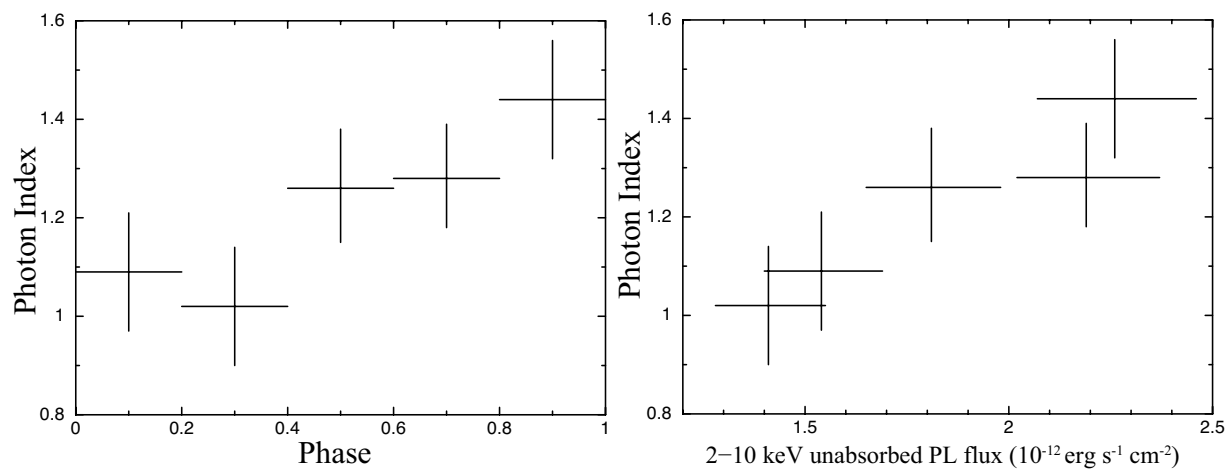}
\end{center}
\caption{(left) Photon index versus pulse phase. The temperature of the BB component is fixed in the fitting. The vertical error bars denote single-parameter 90\% confidence level. (right) Photon index versus 2--10 keV unabsorbed PL flux. Both vertical and horizontal error bars denote single-parameter 90\% confidence level.}
\label{photonindexplot}
\end{figure*}

\red{To explore the counterpart of the pulse profile difference among energy bands} (confirmed in \red{Section} \ref{section:timing/pulseprofiles}), we extracted phase-resolved spectra and conducted spectral fittings to investigate the differences in the spectrum among different pulse phases. Only pn and FPMA data were used, but we have confirmed that there is no significant change in results when we include MOS data in the analysis. %
The definition of the phase is the same as that employed in the analysis of the pulse profiles (\red{Section} \ref{section:timing/pulseprofiles}). We first divided the whole spectrum into two pieces, namely, phases 0.0--0.5 and 0.5--1.0. In addition, we divided the whole spectrum into five pieces. Throughout the entire analysis, the parameter $N_{\rm H}$ was fixed, and either $kT$ or $\Gamma$ was also fixed, adopting the value obtained in \red{Section} \ref{section:spectrum/combined}.\\
\indent The best-fit parameters of each fitting are presented in Table \ref{phaseresolved_bestfitparameters}. All 14 fittings show acceptable values of $\chi^{2}/({\rm d.o.f})$. When we fix $kT$, we clearly see that the photon index $\Gamma$ varies among phases beyond their error ranges. Similarly, when we fix $\Gamma$, we see that the surface temperature $kT$ varies among phases beyond their error ranges. \red{These properties are presented in Figure \ref{BBtempplot} and \ref{photonindexplot}.} These results show that the spectrum changes with the pulse phase, which corresponds to the differences in pulse profiles among energies.

\begin{table*}
  \tbl{ Best-fit parameters of the phase-resolved spectral analysis.\footnotemark[$*$] }{%
  \begin{tabular}{cccccc}
      \hline
      Phase\footnotemark[$\dag$] &  \begin{tabular}{c}$kT$\footnotemark[$\ddag$]\\(keV)\end{tabular} & BB norm & $\Gamma$ & PL norm\footnotemark[$\dag$] & $\chi^{2}_{\nu}$ (d.o.f.)\\ 
      \hline
      0.0--0.5 & $0.52$ (fixed) & \red{$4.7\pm0.2$} & $1.13^{+0.08}_{-0.07}$ & $1.57^{+0.10}_{-0.09}$ & 1.22 (112)\\
      0.5--1.0 & $0.52$ (fixed) & \red{$4.9\pm0.3$} & $1.34\pm0.07$ & $2.13^{+0.12}_{-0.11}$ & 1.22 (127)\\
      0.0--0.2 & $0.52$ (fixed) & \red{$4.5\pm0.4$} & $1.09\pm0.12$ & $1.54^{+0.15}_{-0.14}$ & 1.43 (43)\\
      0.2--0.4 & $0.52$ (fixed) & \red{$4.6\pm0.4$} & $1.02\pm0.12$ & $1.41^{+0.14}_{-0.13}$ & 1.18 (41)\\
      0.4--0.6 & $0.52$ (fixed) & \red{$5.6\pm0.4$} & $1.26^{+0.12}_{-0.11}$ & $1.81^{+0.17}_{-0.16}$ & 0.93 (49)\\
      0.6--0.8 & $0.52$ (fixed) & \red{$5.2^{+0.4}_{-0.5}$} & $1.28^{+0.11}_{-0.10}$ & $2.19^{+0.18}_{-0.17}$ & 1.25 (53)\\
      0.8--1.0 & $0.52$ (fixed) & \red{$4.5\pm0.5$} & $1.44\pm0.12$ & $2.26^{+0.20}_{-0.19}$ & 0.85 (48)\\
       \hline
       0.0--0.5 & $0.49^{+0.02}_{-0.01}$ & \red{$4.8^{+0.2}_{-0.3}$} & $1.21$ (fixed) & $1.68\pm0.06$ & 1.16 (112)\\
       0.5--1.0 & $0.54^{+0.02}_{-0.01}$ & \red{$5.1\pm0.2$} & $1.21$ (fixed) & $1.93\pm0.07$ & 1.24 (127)\\
       0.0--0.2 & $0.50\pm0.03$ & \red{$4.5^{+0.3}_{-0.4}$} & $1.21$ (fixed) & $1.69\pm0.10$ & 1.44 (43)\\
       0.2--0.4 & $0.47^{+0.03}_{-0.02}$ & \red{$4.6\pm0.4$} & $1.21$ (fixed) & $1.64\pm0.10$ & 1.11 (41)\\
       0.4--0.6 & $0.51\pm0.02$ & \red{$5.8\pm0.4$} & $1.21$ (fixed) & $1.78\pm0.10$ & 0.92 (49)\\
       0.6--0.8 & $0.54^{+0.02}_{-0.03}$ & \red{$5.2\pm0.4$} & $1.21$ (fixed) & $2.07\pm0.11$ & 1.24 (53)\\
       0.8--1.0 & $0.56\pm0.03$ & \red{$4.9\pm0.4$} & $1.21$ (fixed) & $1.88\pm0.11$ & 0.94 (48)\\
       \hline
    \end{tabular}}\label{phaseresolved_bestfitparameters}
\begin{tabnote}
\footnotemark[$*$] Only pn and FPMA data are used. Errors denote single-parameter 90\% confidence level. The explanations of the parameters are the same as in Table \ref{bestfitparameters}. \\ 
\footnotemark[$\dag$] The offset of the phase is set in accordance with the pulse profiles (Figures \ref{pulseprofiles} and \ref{pulseprofile_energy}).\\
\footnotemark[$\ddag$] When the parameters are fixed, we adopt the value of the best-fit model of the joint fitting \red{of the average spectrum} with XMM-Newton and NuSTAR (see Table \ref{bestfitparameters}). \red{$N_{\rm H}$ is always fixed to $1.96\times10^{22}\;{\rm cm^{-2}}$.}\\
\end{tabnote}
\end{table*}

\section{Discussion}
\subsection{Broad-band spectra of SGR\,1900+14}\label{discussion/spectrum}
Our simultaneous broad-band observation with XMM-Newton and NuSTAR successfully measured the spectrum of SGR\,1900+14 much more precisely than previous observations. The obtained spectral properties can be compared with \red{those obtained in} previous observations performed in the quiescent stages of the object, where bursting activities were not seen.\\
\indent Table \ref{spectrum_parameters_comparison} presents a comparison of our spectral analysis results with the previous XMM-Newton and Suzaku results in 2005, 2006, and 2009 \citep{Mereghetti2006, Enoto2017}. We found a clear decrease in the flux of SGR\,1900+14, which can be attributed to a continuous decline since the giant flare. We confirm that the 1-10 keV flux $F_{1-10}$ decreases by $\sim25$--$40\%$. The 15--60 keV flux $F_{15-60}$ also shows a possible decrease, but this cannot be confirmed because of background model uncertainties of Suzaku HXD \citep{Fukazawa2009}: when we employed the results of another analysis of the same data set \citep{Enoto2010}, the decrease of $F_{15-60}$ was not found.\\
\indent Although the flux shows a clear decrease, the spectral shape does not show significant changes from \red{those} observed in 2005, 2006, and 2009. The obtained $N_{\rm H}$ and $kT$ values are perfectly matched with those of Suzaku results; they are also comparable with those \red{obtained} from the XMM-Newton observations. As for the photon index $\Gamma$, our results do not agree with previous studies in terms of XMM-Newton or Suzaku, which is presumably due to the absence of the hard X-ray observation for XMM-Newton and uncertainties of the background model of Suzaku HXD \citep{Fukazawa2009}. Another analysis performed on the same Suzaku observation gives $\Gamma=1.2(5)$ and $1.4(3)$ for the 2006 and 2009 data \citep{Enoto2010}, respectively, both of which agreeing with our result within the errors. Therefore, we suggest that SGR\,1900+14 has been quiescent for more than 10 years.\\
\indent In addition, we found a clear decrease in \red{the 1--60 keV unabsorbed soft component (BB) flux} $F_{\rm s}$ over 10 years. We also found a trend where the PL component may decrease even faster than the BB component because the hardness ratios, which are defined in two ways as $\eta=F_{15-60}/F_{1-10}$ or $\xi=F_{\rm h}/F_{\rm s}$, show decreases, \red{where $F_{\rm h}$ denotes the 1--60 keV unabsorbed PL flux.} However, this result cannot be confirmed because the evaluation of the hardness ratios largely depends on the uncertainties of the HXD background model\citep{Fukazawa2009}. %Therefore, we suggest a trend that the PL component may decrease even faster than the BB component, although the confirmation is beyond our scope.

\begin{table*}
  \tbl{ Comparison of spectral analyses of SGR\,1900+14. }{%
  \scalebox{0.85}{
  \begin{tabular}{cccccccccc}
      \hline
      \begin{tabular}{c}Satellite\\Date\end{tabular} & \begin{tabular}{c}$N_{\rm H}$\\$(10^{22}\;{\rm cm^{-2}})$\end{tabular} & \begin{tabular}{c}$kT$\\(keV)\end{tabular} & $\Gamma$ & $F_{1-10}$\footnotemark[$*$] & $F_{15-60}$\footnotemark[$*$] & \begin{tabular}{c}Abs. HR\footnotemark[$\dag$]\\$\left(\eta=\frac{F_{15-60}}{F_{1-10}}\right)$\end{tabular} & $F_{\rm s}$\footnotemark[$\ddag$] & $F_{\rm h}$\footnotemark[$\ddag$] & \begin{tabular}{c}HR\footnotemark[$\dag$]\\$\left(\xi=\frac{F_{\rm h}}{F_{\rm s}}\right)$\end{tabular}\\
       \hline
       \begin{tabular}{c}XMM-Newton\footnotemark[$\S$]\\2005-09-20\end{tabular} & $2.12\pm0.08$ & $0.47\pm0.02$ & $1.9\pm0.1$ & $4.3\pm0.2$ & $-$ & $-$ & $-$ & $-$ & $-$\\
       \begin{tabular}{c}XMM-Newton\footnotemark[$\S$]\\2006-04-01\end{tabular} & $2.3^{+0.1}_{-0.2}$ & $0.47\pm0.03$ & $1.9\pm0.2$ & $4.8\pm0.2$ & $-$ & $-$ & $-$ & $-$ & $-$\\
       \begin{tabular}{c}Suzaku\footnotemark[$\|$]\\2006-04-01\end{tabular} & $1.8\pm0.3$ & $0.57\pm0.2$ & $0.96\pm0.14$ & $5.3\pm0.5$ & $20.6\pm5.5$ & $3.9\pm1.1$ & $4.6^{+0.1}_{-0.6}$ & $25.0^{+3.2}_{-3.4}$ & $5.4^{+0.7}_{-1.0}$\\
       \begin{tabular}{c}Suzaku\footnotemark[$\|$]\\2009-04-26\end{tabular} & $1.9\pm0.1$ & $0.52\pm0.02$ & $0.78\pm0.09$ & $4.3\pm0.1$ & $16.5\pm3.5$ & $3.8\pm0.8$ & $4.5\pm0.3$ & $26.3^{+2.9}_{-2.5}$ & $5.9\pm0.7$\\
       \begin{tabular}{c}XMM+NuSTAR\\2016-10-20\end{tabular} & $1.96\pm0.11$ & $0.52\pm0.02$ & $1.21\pm0.06$ & $3.11\pm0.03$ & $6.8\pm0.3$ & $2.2\pm0.1$ & \red{$3.4\pm0.2$} & \red{$9.8^{+0.3}_{-0.7}$} & $2.9^{+0.2}_{-0.3}$\\
       \hline
    \end{tabular}}}\label{spectrum_parameters_comparison}
\begin{tabnote}
\footnotemark[$*$] Absorbed flux \red{in the 1--10 keV and 15--60 keV energy bands} in units of $10^{-12}\;{\rm erg\;s^{-1}\;cm^{-2}}$.\\ 
\footnotemark[$\dag$] HR denotes the hardness ratio.\\
\footnotemark[$\ddag$] $F_{\rm s}$ and $F_{\rm h}$ denote 1--60 keV unabsorbed BB and PL flux, respectively, in units of $10^{-12}\;{\rm erg\;s^{-1}\;cm^{-2}}$.\\
\footnotemark[$\S$] The results of analyses are derived from \citet{Mereghetti2006}. Because the observation is confined to the soft X-ray range, we cannot obtain information from the hard X-ray range.\\ 
\footnotemark[$\|$] The results of analyses are derived from \citet{Enoto2017}.\\
\end{tabnote}
\end{table*}

\subsection{Timing properties of SGR\,1900+14}
\subsubsection{Rotation period and its evolution}
\begin{figure}
\begin{center}
\includegraphics[width=80mm]{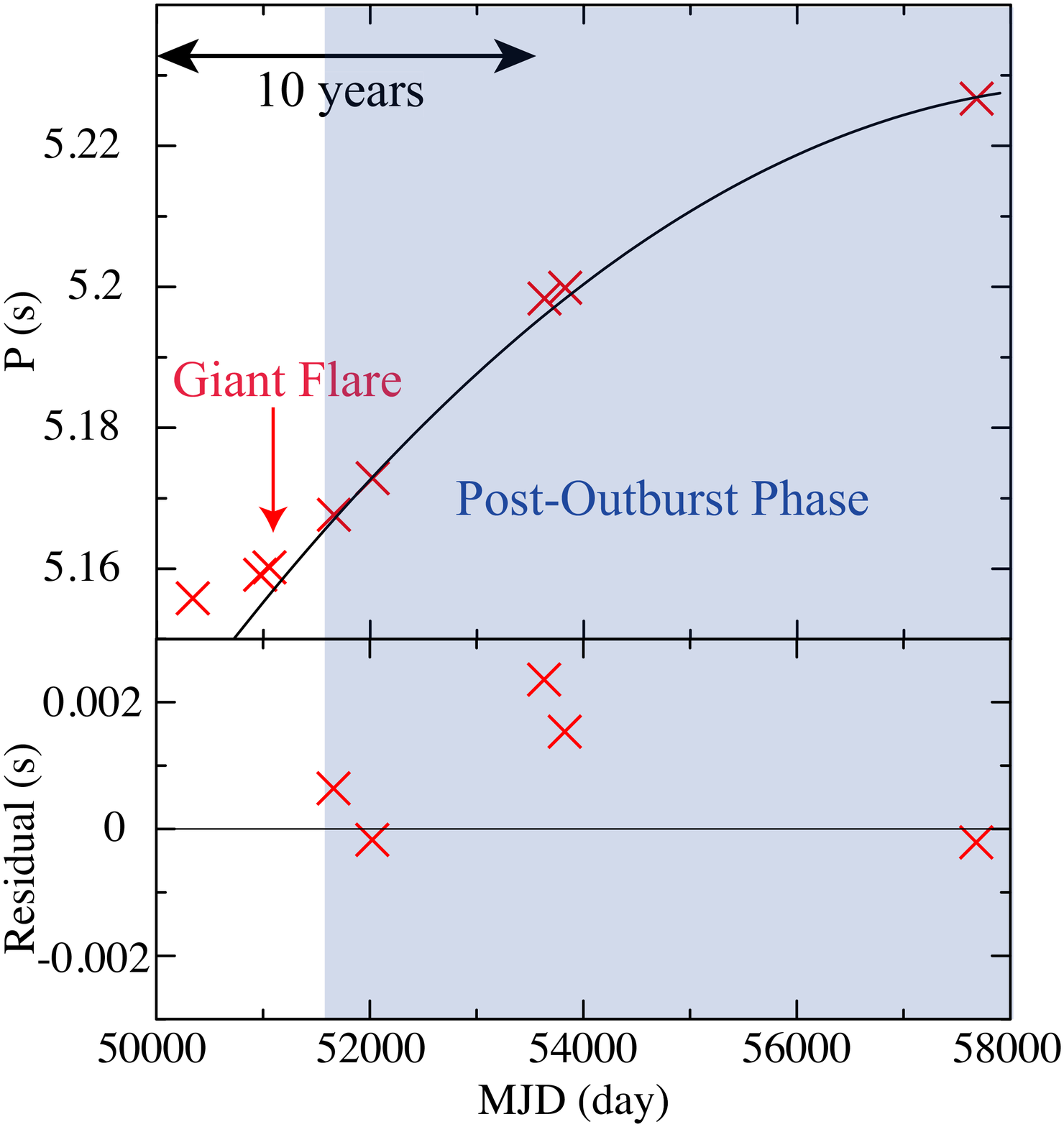}
\end{center}
\caption{Long-term evolution of the rotation period of SGR\,1900+14. Red crosses denote the rotation period of SGR\,1900+14 at each time, among which the right one is obtained from our work. Other data were obtained from previous studies \citep{Mereghetti2006, Marsden1999, Kouveliotou1999, Woods2002, Woods2003}. Errors are much smaller than the \red{crosses}. The red arrow denotes \red{the epoch of the} giant flare (MJD=51052). The blue hatched area denotes the post-outburst phase, which we defined in this work (see text for details), covering the data \red{since} MJD=51660. The black curve denotes the quadratic model fitted to the data in the post-outburst phase.}\label{longtermevolution_figure}
\end{figure}

The detection of coherent pulsation (\red{Section} \ref{section:timing/coherentpulsation}) indicates that the rotation period of SGR\,1900+14 was $5.22669(3)\;{\rm s}$ on 2016 October 20 to 23. Figure \ref{longtermevolution_figure} and Table \ref{longtermevolution_table} present the long-term evolution of the rotation period of SGR\,1900+14, where we see a monotonic spin-down \red{over the past} 20 years. As shown in Figure \ref{longtermevolution_figure}, the spin-down rate $\dot{P}$ has a large fluctuation. This behavior is typical for magnetars (e.g., CXOU\,J171405.7-381031: \cite{Halpern2010}; Swift\,J1822.3-1606: \cite{Tong2013}) and is in clear contrast with the behavior of normal radio pulsars (e.g., the Crab pulsar: \cite{Terada2008}).\\
\indent Here, we define the ``post-outburst phase'' of SGR\,1900+14 as the hatched area in Figure \ref{longtermevolution_figure}. In this phase, the unstable fluctuations of $\dot{P}$ soon after the giant flare \citep{Woods2002} cease and $\dot{P}$ decreases monotonically. Assuming a constant decay of $\dot{P}$, we determined a quadratic function (black curve in Figure \ref{longtermevolution_figure}) that describes the trend of $\dot{P}$ in the post-outburst phase. We obtained $\ddot{P}=-3.1\times10^{-19}\;{\rm s^{-1}}$, which suggests a monotonic decrease in the spin-down rate. $\dot{P}$ shows a drastic decrease from $2.0\times10^{-10}\;{\rm s\;s^{-1}}$ in 2000 April (MJD=51660) to $3.3\times10^{-11}\;{\rm s\;s^{-1}}$ in 2016 October (MJD=57681).\\% Note that we did not calculate the errors of the fitting because the errors of data are too small and the fitting is not accurate to be exact.\\
\indent The trend of $\dot{P}$ in the post-outburst phase could be explained by the decrease in the toroidal component (or decrease in the twist) of the magnetic fields in the magnetosphere (twisted magnetosphere model: \cite{Thompson2002, Beloborodov2009}), which is considered to cause a decrease in the spin-down torque. Since the spin-down torque is proportional to the spin-down rate $\dot{P}$, the observed monotonic decrease in $\dot{P}$ during the post-outburst phase can imply that the twist of the magnetic fields has declined monotonically for more than 15 years since it reached its maximum, which was soon after the giant flare.\\%, which may have a causal relation with the giant flare activities.\\
\indent Similar behaviors of other magnetars have been reported. For example, PSR\,J1622-4950, which entered its outburst in or before 2007 June, showed an unstable fluctuation of the spin-down rate soon after its outburst and then a monotonic decrease with a constant rate \citep{Scholz2017}. XTE\,J1810-197, which is one of the few magnetars that display radio emissions, experienced an outburst in 2003 August along with a subsequent large fluctuation of $\dot{P}$, which was followed by a slow and gradual decrease of $\dot{P}$ \citep{Pintore2016}. Another example is SGR\,J1745-2900, which also showed an unstable behavior of $\dot{P}$ soon after its outburst in 2013 April and subsequent gradual monotonic decrease of $\dot{P}$ \red{\citep{CotiZelati2015, CotiZelati2017}}. The values of $\ddot{P}$ in the  post-outburst phases are $-2.0\times10^{-20}\;{\rm s^{-1}}$, $-5.5\times10^{-21}\;{\rm s^{-1}}$, and $-1.8\times10^{-19}\;{\rm s^{-1}}$ for PSR\,J1622-4950, XTE\,J1810-197, and SGR\,J1745-2900, respectively. All three values of $\ddot{P}$ are negative, as is that for SGR\,1900+14, whose $|\ddot{P}|=3.1\times10^{-19}\;{\rm s^{-1}}$ is larger than those of \red{the three above-mentioned examples}. We point out the possibility that the decay of the twist of magnetic fields in the magnetosphere of SGR\,1900+14 has been lasting for more than 15 years, which is much longer than the duration reported for other magnetars. This may be due to the scale of the outburst; we call the outburst of SGR\,1900+14 a giant flare with a peak luminosity of $\gtrsim10^{44}\;{\rm erg\;s^{-1}}$, while other magnetars that have such an outburst have peak luminosities of $\gtrsim10^{35}\;{\rm erg\;s^{-1}}$ \citep{Anderson2012, Gotthelf2004, CotiZelati2015}. Still, another magnetar SGR 1806-20, which experienced a giant flare and is often regarded as a set with SGR 1900+14, was reported to have a similar trend of the timing evolution \citep{Younes2015, Younes2017}. It should be noted that since observations have not been conducted so frequently for SGR\,1900+14, this is not necessarily the case. For example, the monotonic decrease of $\dot{P}$ may have ceased at some point between 2006 and 2016, where we have no data of the rotation period, and thereafter began linear spin-down behavior like normal radio pulsars.\\
\indent Employing the newly obtained $P$ and $\dot{P}$, we can evaluate two important parameters: the dipole magnetic field $B_{\rm d}$ and the characteristic age $\tau_{\rm c}$ of SGR\,1900+14. These are defined as
\begin{eqnarray}
B_{\rm d}&=&3.2\times10^{19}\left(\frac{P}{\rm s}\cdot\frac{\dot{P}}{\rm s\;s^{-1}}\right)^{1/2}\;{\rm G},\\
\tau_{\rm c}&=&\frac{(P/{\rm s})}{2(\dot{P}/{\rm s\;s^{-1}})}.
\end{eqnarray}
The $P$ and $\dot{P}$ values on 2016 October 20 give $B_{\rm d}=4.3\times10^{14}\;{\rm G}$ and $\tau_{\rm c}=2.4\;{\rm kyr}$, respectively. This result means that SGR\,1900+14 is older and has a smaller dipole magnetic field than previously reported ($B_{\rm d}=7\times10^{14}\;{\rm G}$ and $\tau_{\rm c}=0.9\;{\rm kyr}$ reported by \cite{Olausen2014} and based on \citet{Mereghetti2006}). This is because the previous studies determined these parameters over a short duration without considering the long-term evolution of the rotation period.

\begin{table*}
  \tbl{ List of rotation periods of SGR\,1900+14.}{%
  \begin{tabular}{ccccc}
      \hline
      \begin{tabular}{c}Date\\(YYYY-MM-DD)\end{tabular} & Telescope & \begin{tabular}{c}Exposure\footnotemark[$*$]\\(ks)\end{tabular} & \begin{tabular}{c}$P$\\(s)\end{tabular} & Reference\\
      \hline
      1996-09-04--1996-09-18 & RXTE & 47 & $5.1558199(29)$ & \cite{Marsden1999}\\
      1998-05-31--1998-06-09 & RXTE & 41.7 & $5.159142(3)$ & \cite{Kouveliotou1999}\\
      1998-08-28 & RXTE & 2.5 & $5.160199(2)$ & \cite{Kouveliotou1999}\\
      2000-04-26 & RXTE & 10 & $5.16765(3)$ & \cite{Woods2002}\\
      2001-04-18--2001-05-05 &RXTE & 128 & $5.17284827(40)$ & \cite{Woods2003} \\
      2005-09-20 & XMM-Newton & 47.4 & $5.198346(3)$ & \cite{Mereghetti2006}\\
      2006-04-01 & XMM-Newton & 15.7 & $5.19987(7)$ & \cite{Mereghetti2006}\\
      2016-10-20--2016-10-23 & NuSTAR & 123 & $5.22669(3)$ & This work\\
      \hline
    \end{tabular}}\label{longtermevolution_table}
\begin{tabnote}
\footnotemark[$*$]  If there is a difference in exposure times among detectors, the maximum of them is cited.\\ 
%\footnotemark[$\dag$] 
%\footnotemark[$\ddag$] 
%\footnotemark[$\S$] 
\end{tabnote}
\end{table*}

\subsubsection{Pulse profiles}
We found no significant variation in the pulse fraction over 10 years. The pulse profiles obtained with XMM-Newton and NuSTAR all give pulse fractions of 15--20\%, which agrees with the previous XMM-Newton observations in 2005 and 2006 \citep{Mereghetti2006}.\\%, also giving pulse fractions of 15--20\% under 10 keV.\\
%\indent The energy-dependent pulse profiles show that \red{3--10 keV is almost sinusoidal, while 10--20 keV is not.} 
\indent \redd{The pulse profile shape is almost sinusoidal in the 3--10 keV energy band, while it is more structured in the 10--20 keV.} Since the power-law component is dominant in the range 10--20 keV (see Figure \ref{combinedspectra}), we may see contributions of \red{higher multipolar components of the magnetic field} on the stellar surface.
%\indent The energy-dependent pulse profiles show an interesting trend that 3--10 keV and 10--20 keV profiles differ in morphology, the former almost sinusoidal while the latter not. Since the power-law (PL) component is prominent above 10 keV (see figure \ref{combinedspectra}), this could be interpreted as a characteristic behavior of PL component which the blackbody (BB) component does not show. However, while the boundary of the morphology of the pulse profiles is around 10 keV, the threshold energy at which the dominant component of the spectrum changes from BB to PL is around 5 keV, according to our spectral analysis (see sub-subsection \ref{section:spectrum/combined} and figure \ref{combinedspectra}). What causes this disagreement is still unclear and is beyond the scope of this paper.

\subsection{Phase-dependent fluctuation of power-law component}
Our phase-resolved spectral analysis (\red{Section} \ref{section:phasespectra}) suggests that the spectral shape varies with the pulse phase. Figure \ref{BBtempplot} shows the BB \red{temperature} as a function of the pulse phase and the relation between the BB temperature and the BB flux, which was obtained with a fixed PL photon index. Although the BB temperature shows variations with the pulse phase, we did not find its correlation with the BB flux, as shown in Figure \ref{BBtempplot}.\\
\indent \red{As a next step, we explored how the PL component varies with the pulse phase.} Figure \ref{photonindexplot} shows the photon index as a function of the pulse phase and the relation between the photon index and the unabsorbed PL flux. We found a positive correlation between the photon index and the unabsorbed PL flux. Note that although these two parameters can be \red{covariant and} positively coupled, we confirmed that the positive correlation is significant by checking their error contours. The relation between the photon index and the PL flux is consistent with the trend of the systematic analysis of various magnetars \citep{Enoto2010, Enoto2017}. \red{They reported a signature that the hard PL component above 10 keV shows softer spectral photon index as dipole magnetic fields of a magnetar become stronger. This trend was interpreted as a process that the higher magnetic field leads to more photon splittings into lower energy photon, i.e., the softer power-law spectrum.}\\
\indent The positive correlation between the photon index and the PL flux means that the spectrum gets softer when the emitting area, which is usually a hot spot, is better oriented to the observer. \red{We can interpret this trend in terms of photon splitting, which is a nonlinear effect of QED under extremely strong magnetic fields \citep{Baring1998, Baring2001, Chistyakov2012}.} When the PL flux is at the pulse maximum, we can see the region with the strongest magnetic fields, which cause more photon splittings and thus a softer spectrum. Another possibility is that we may see the difference of the path length of splitting photons: the photons from the direction of the magnetic pole travel longer in magnetosphere and experience more splittings. In both cases, our results are consistent with the trend of the photon splitting model. 

%Previous studies on various magnetars \citep{Enoto2010, Enoto2017} have suggested that because a higher magnetic field leads to more photon splittings, it also leads to a softer power-law spectrum because more photons split into lower energy photons. This hypothesis can be applied to our observation.

\section{Conclusions}
\red{We performed the first simultaneous broad-band observations} of the magnetar SGR\,1900+14 covering 0.1--78 keV, making full use of the high sensitivity of XMM-Newton and NuSTAR. The NuSTAR hard X-ray coverage enabled us to detect the source up to 70 keV, with a 60--70 keV source significance of $6.5\sigma$.\\% It has allowed us to obtain data from the object in 0.1--70 keV and conduct a detailed analysis with little background contamination.\\
\indent \red{The spectrum of SGR\,1900+14 was well fitted by a typical magnetar spectral model: BB plus PL.} We have successfully determined the properties of the spectrum with a higher accuracy than any previous studies, especially for the hard-tail power-law component. The obtained parameters are $N_{\rm H}=(1.96\pm0.11)\times10^{22}\;{\rm cm^{-2}}$, $kT=0.52\pm0.02\;{\rm keV}$, and $\Gamma=1.21\pm0.06$. We have found that the flux has decreased for more than 10 years, presumably because of the decline from the giant flare, while the spectral shape exhibited no significant variations.\\
\indent Timing analysis allowed us to determine that the rotation period of SGR\,1900+14 on 2016 October 20 to 23 was $5.22669(3)\;{\rm s}$. The long-term evolution of the rotation period shows a monotonic decrease in $\dot{P}$ in the post-outburst phase, which suggests that the twist of the magnetic fields in the magnetosphere has been decaying for more than 15 years. Its pulse fraction was in the range 15--20\%, showing no variation in the past 10 years. The energy-dependent pulse profiles present an interesting trend that the 3--10 keV band is almost perfectly sinusoidal, while the 10--20 keV band contains higher harmonics.\\%, which is the evidence of the characteristic behavior of the PL component.\\
\indent Combining the spectral and temporal analyses, we succeeded in obtaining the phase-resolved spectra. It shows an interesting feature that the photon index and unabsorbed PL flux have a positive correlation, which suggests that we may see differences in the process of photon splitting within one phase \red{cycle}. 

\begin{ack}
We thank the anonymous referee for his/her valuable comments. We thank Shinpei Shibata, Kuniaki Masai, Hiromasa Suzuki, Takahiro Matsumoto, and Kazuo Makishima for their helpful advice and discussions. This work was also supported by the Grant-in-Aid for Scientific Research on Innovative Areas ``Toward new frontiers: Encounter and synergy of state-of-the-art astronomical detectors and exotic quantum beams'' (18H05459; AB). We acknowledge support from JSPS/MEXT KAKENHI grant numbers 18H05459 (AB), 19K03908 (AB), 18H05861 (HO), 16H03954 (HO), 15H00845 (TE), 17K18776 (TE), and 18H04584 (TE). 
\end{ack}

\end{document}